\documentclass{aa}
\usepackage{times}

\newcommand{\ltsima} {$\; \buildrel < \over \sim \;$}
\newcommand{\simlt}  {\lower.5ex\hbox{\ltsima}}            % < over MMM
\newcommand{\gtsima} {$\; \buildrel > \over \sim \;$}
\newcommand{\simgt}  {\lower.5ex\hbox{\gtsima}}            % > over MMM
\newcommand{\ferg}{erg cm$^{-2}$ s$^{-1}$ }

\newcommand{\be} {\begin{equation}}

\newcommand{\ee} {\end{equation}}

\newcommand{\BSAX}{{\em Beppo}SAX}
\newcommand{\RXTE}{{\em Rossi}XTE}
\newcommand{\xmm}{{\em XMM-Newton}}
\newcommand{\bc}{\begin{center}}
\newcommand{\ec}{\end{center}}

\def \nh {N${\rm _H}$}
\def \hcm {\hbox {\ifmmode $ atoms cm$^{-2}\else atoms cm$^{-2}$\fi}}

\def\ee {1E~1048.1--5937~}
\def\sgr {SGR~1806--20~}
\def\SGR {SGR~1900+14~}

\begin{document}

\title
{The calm after the storm: \xmm\ observation of \sgr\ two months
after the Giant Flare of 2004 December 27\thanks{Based on
observations obtained with \xmm, an ESA science mission with
instruments and contributions directly funded by ESA Member States
and NASA} }

\author{ A.~Tiengo\inst{1,2}, P.~Esposito\inst{1} ,
S.~Mereghetti\inst{1}, N.~Rea\inst{3,4}, L.~Stella\inst{3},
G.L.~Israel\inst{3},
 R.~Turolla\inst{5},
S.~Zane\inst{6} }

\institute { {1) INAF,  Istituto di Astrofisica Spaziale e Fisica
Cosmica, Sezione di Milano ``G.Occhialini'', via Bassini 15, I-20133
Milano, Italy}
\\
{2) Universit\`a degli Studi di Milano, Dipartimento di Fisica, via
Celoria 16, I-20133 Milano, Italy}
\\
 {3) INAF, Osservatorio Astronomico di Roma, via
Frascati 33, I-00040 Monteporzio Catone, Roma, Italy}
\\
 {4) SRON - National Institute for Space Research, Sorbonnelaan, 2, 3584 CA,
Utrecht, The Netherlands}
\\
{5) Universit\`a di Padova, Dipartimento di Fisica, via Marzolo 8,
I-35131 Padova, Italy}
 \\
{6) Mullard Space Science Laboratory, University College London,
Holmbury St. Mary, Dorking Surrey, RH5 6NT, UK}
 }

\offprints{A. Tiengo, tiengo@mi.iasf.cnr.it}

\date{Received / Accepted}

\authorrunning{Tiengo et al. }
\titlerunning{ \sgr}
%\maketitle

\abstract{
%The first \xmm\ observation of \sgr\ after its giant flare
%was performed 70 days after the event. A comparison with the
%previous \xmm\ observations shows that the pulsed fraction and
%spin-down rates have significantly decreased and the spectrum
%slightly softened. These changes may indicate a global
%reconfiguration of the neutron star magnetosphere. The presence of a
%blackbody component in addition to an absorbed power-law model is
%strongly required to fit the X-ray spectrum. All the \xmm\ spectra
%of \sgr\ are consistent with a power-law component with variable
%intensity and rather stable slope and a constant blackbody component
%contributing only to a small fraction of the observed flux.
\xmm\ observed the soft gamma repeater \sgr\ about two months after
its 2004 December 27 giant flare. A comparison with the previous
observations taken with the same instrument in 2003--2004 shows that
the pulsed fraction and the spin-down rate have significantly
decreased and that the spectrum slightly softened. These changes may
indicate a global reconfiguration of the neutron star magnetosphere.
The spectral analysis confirms that the presence of a blackbody
component in addition to the power-law is required. Since this
additional component is consistent with being constant with respect
to the earlier observations, we explore the possibility of
describing the long-term spectral evolution as only due to the
power-law variations. In this case, the slope of the power-law does
not significantly change and the spectral softening following the
giant flare is caused by the increase of the relative contribution
of the blackbody over the power-law component.  }

\maketitle

\section{Introduction}

On 2004 December 27 most of the X-ray and $\gamma$--ray satellite
instruments were saturated by the brightest extra-solar event ever
recorded (Borkowski et al. 2004, Hurley et al. 2005, Palmer et al.
2005a, Terasawa et al. 2005). Its initial $\sim$0.2 s long spike was
bright enough to cause significant perturbation in the Earth's
ionosphere (Campbell et al. 2005) and to be detected also through
its reflection on the Moon's surface (Mazets et al. 2005, Mereghetti
et al. 2005a). The presence of a pulsating tail modulated at the
spin period of the Soft Gamma Repeater (SGR) \sgr\ allowed it to be
identified as a giant flare from this source, in analogy to similar
(but less energetic) events registered from SGR~0526--66 and \SGR\
on 1979 March 5 (Mazets et al. 1979) and 1998 August 27 (Hurley et
al. 1999), respectively.
%The identification with \sgr\ was then finally
%confirmed by the appearance of a new bright radio source at the SGR
%location (Gaensler et al. 2005, Cameron et al. 2005).

%The SGR giant flares can be interpreted in the framework of the
%magnetar model
Giant flares are the most spectacular manifestations of a small
class of young neutron stars which are thought to be magnetars. A
magnetar is a neutron star with a very strong magnetic field
(B$\sim$10$^{14}$--10$^{15}$ G), whose decay powers its high energy
emission (Duncan \& Thompson 1992, Paczynski 1992, Thompson \&
Duncan 1995). Apart from the very rare giant flares, SGRs frequently
emit short ($\sim$0.1~s) bursts of soft $\gamma$-rays and are
persistent sources of X-rays. This emission is pulsed, and the
periods (in the $\sim$5--10~s range) steadily increase at a rate of
$\sim$10$^{-10}$--10$^{-11}$ s s$^{-1}$. These properties are shared
with the Anomalous X-ray Pulsars (AXPs), which are also thought to
be magnetars (Thompson \& Duncan 1996, see Woods \& Thompson 2004
for a recent review).

%The magnetar model was developed to explain the bursting emission of
%SGRs, but was also able to interpret their persistent X-ray emission
%and that of the Anomalous X-ray Pulsars (AXPs).

Here we present the results of an \xmm\ target of opportunity
observation of \sgr, the first one performed with this satellite
after the giant flare.
% and a comparison with the source status
%observed during four \xmm\ observations performed between April 2003
%and October 2004 (Mereghetti et al. 2005b).

\section{Observation and data analysis}

Due to visibility constraints, \xmm\ could not observe \sgr\ before
March 2005. The observation was performed on 2005 March 7 and had a
duration of 25 ks. An instrumental configuration similar to the
previous observations of this source (Mereghetti et al. 2005b) was
kept: the EPIC PN (Str{\"u}der et al. 2001) was operated in Small
Window mode (time resolution 6 ms) while the EPIC MOS (Turner et al.
2001) had the MOS1 unit in Timing mode (time resolution 1.5 ms) and
the MOS2 in Full Frame mode (time resolution 2.6 s). Both the PN and
MOS mounted the medium thickness filter.

The data were processed using Version 6.1.0 of the \xmm\
\emph{Science Analysis System} (SAS)
%\footnote{\texttt{http://xmm.vilspa.esa.es/external/xmm\_sw\_cal/sas\_frame.shtml}})
and the most recent calibration files (last update on 2005 May 14).

\begin{figure*}[htbp!]
\vspace{8.cm} \includegraphics{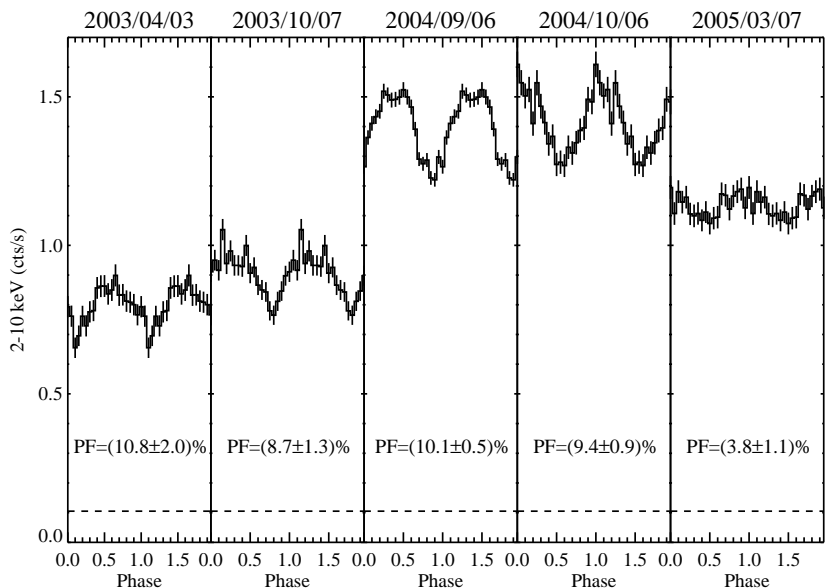} \caption{Background subtracted pulse
profiles in the 2--10 keV energy range of \sgr\ as seen by the PN
camera during the five \xmm\ observations (the dates of the
observations are reported at the top of each panel). The dashed line
shows the blackbody contribution to the phase averaged count rate
when the source spectrum is fitted using an absorbed power-law plus
blackbody, with the absorption and blackbody parameters linked to
the same value
%only the power-law parameters free to vary independently
during the five observations (last line in Table 1). The pulsed
fractions (PF) computed from a sinusoidal fit of the profiles are
also indicated.}
\end{figure*}

The MOS2 data allowed us to search for extended emission over the
whole field of view (15$'$ radius), but apart from a bright point
source at the position of \sgr\ and a couple of weaker point
sources, no other X-ray emitting structures were detected through a
visual check of images in different energy bands. Given the huge
flux of X-ray photons emitted during the giant flare and the large
amount of dust very likely present along the line of sight, a dust
scattering echo might be expected in X-ray observations following
the giant flare (see e.g. Vaughan et al. 2004). However, at these
late times, independent on the dust spatial distribution and
composition, the scattering angle would be $>20'$ if single
scattering occurs, and the efficiency for X-ray scattering at these
large angles is very small (Draine 2003). The probability of
multiple scattering at smaller angles is also rather small and
therefore no information can be obtained from the lack of an X--ray
dust echo of the \sgr\ giant flare.

In the following we report the results of the timing and spectral
analysis of the PN data. A similar analysis was performed also on
MOS data giving consistent results.

The time of arrival of the detected events were corrected to the
Solar System barycenter and the source and background photons
(events with pattern 0--4) were extracted from circular regions of
40$''$ radius. The careful analysis of the source and background
lightcurve led to the discovery of two weak bursts ($\sim$10 counts
each) of the duration of $\sim$0.1 s.

From folding and phase fitting analysis of the source lightcurve, a
spin period of 7.5604$\pm$0.0008 s was measured.
%This period is significantly smaller
%than the value obtained extrapolating to the time of this
%observation the spin-down of 5.5$\times$10$^{-10}$ s s$^{-1}$
%derived from previous \xmm\ observations (Mereghetti et al. 2005b).
%This result confirms the lower spin-down already measured in \RXTE\
%observations performed after the giant flare (Woods...).
The background subtracted pulse profile in the 2--10 keV energy band
is shown in Figure 1 together with those of the previous \xmm\
observations (Mereghetti et al. 2005b). Being extracted from the
same instrument and regions, these profiles can be directly
compared: the average count rate is lower than in 2004 but still
higher than in 2003. A sinusoidal fit to the profile gives a pulsed
fraction of (3.8$\pm$1.1)\%, significantly smaller than the
$\sim$10\% value previously observed.

The source spectrum was extracted from a circle of 25$''$ radius
%for PN and MOS2 data and from a stripe ...$''$ wide for the MOS1 data in
%Timing mode, where photon position is kept only along one dimension.
and the background spectrum from composite regions located on the
same chip as the source. We selected events with patterns 0--4
%for the PN, 0 for MOS1, and 0--12 for the MOS2, while
and excluded time intervals with high particle background, obtaining
a net exposure time of 14.7 ks. The spectrum was rebinned to have at
least 30 counts in each energy bin and not to oversample the
instrumental energy resolution. The fits were performed in the
energy range 1.5--12.0 keV, since the source is heavily absorbed and
only few counts are detected at lower energies.

%using the XSPEC v.12 software package \citep{arnaud96}.
A fit with an absorbed power-law gives an unacceptable
%We rejected the fit with an absorbed power law based upon the
%large
$\chi^2$ value, while a good fit can be obtained by adding a
blackbody component (see Table 1).
%The best fit parameters are photon index $\Gamma=$0.8, blackbody
%temperature kT$_{BB}=$0.91 keV,
%and a hydrogen column density of $6 \times 10^{22}$ cm$^{-2}$.
In order to study the spectral evolution of \sgr\ after the giant
flare, we have reprocessed also the data of the previous \xmm\
observations using the most recent calibration files and extracted
the PN spectra similarly to what reported in Mereghetti et al.
(2005b). Then we fitted simultaneously the five spectra with the
power-law plus blackbody model keeping most of the parameters linked
to common values. A very good fit was obtained allowing only the
power-law photon index and normalization to vary independently (see
Table 1). The long term evolution of the power-law parameters
derived from this simultaneous fit is shown in Figure 2, while the
very small contribution of the fixed blackbody to the observed flux
of \sgr\ is indicated in Figure 1.

\begin{table*}[htbp!]
\begin{center}
  \caption{Results of spectral analysis. Errors are at the  90\% c.l. for a single interesting parameter. }
    \begin{tabular}[c]{cccccc}
\hline
\nh\   & $\Gamma$      &  kT$_{BB}$   & R$_{BB}$ & 2--10 keV flux & $\chi^2_{red}$ (d.o.f) \\
 (10$^{22}$ cm$^{-2}$) & & (keV) & (km)$^{(a)}$ & (\ferg)&  \\
\hline
 6.8$^{+0.4}_{-0.1}$ & 1.68$^{+0.10}_{-0.04}$ & --- & ---& 1.91$\times10^{-11}$& 1.49 (72) \\
 6.0$\pm$0.2 & 0.8$^{+0.2}_{-0.3}$ & 0.91$^{+0.14}_{-0.05}$ & 1.9$^{+0.2}_{-0.3}$ & 1.92$\times10^{-11}$& 1.02 (70) \\
 6.6$\pm$0.2$^{(b)}$ & 1.42$^{+0.08}_{-0.07}$$^{(b)}$ & 0.69$^{+0.06}_{-0.07}$$^{(b)}$ & 2.1$^{+0.5}_{-0.9}$$^{(b)}$ & 1.93$\times10^{-11}$$^{(b)}$& 0.96 (358)$^{(b)}$ \\
\hline
\end{tabular}
\end{center}
\begin{small}
$^{(a)}$ Radius at infinity assuming a distance of 15 kpc\\
$^{(b)}$ Simultaneous fit to the five \xmm\ observations with only
the power-law photon index and normalization free to vary independently\\
%\\
\end{small}
\label{spectra}
\end{table*}

%we tried to fit the spectrum keeping most of the parameters fixed to
%the values reported in Mereghetti et al. (2005b) for the best
%quality spectrum (that of the longest observation, performed in
%September 2004), but acceptable fits were obtained only leaving at
%least ... parameters free to vary. For a direct comparison,  A
%simultaneous fit of the five \xmm\ spectra allowing only the
%power-law photon index and normalization to vary independently
%yields a low $\chi^2$ value (see Table 1).

\section{Discussion}

The new data obtained after the giant flare of 2004 December 27 show
significant differences with respect to the previous \xmm\
observations. During the four observations performed from April 2003
to October 2004, \sgr\ had an almost constant pulsed fraction of
$\sim$10\% (Mereghetti et al. 2005b), while 70 days after the giant
flare it had decreased to $\sim$4\%.  This was already noted in the
\emph{Chandra} observation of 2005 February 8 (Rea et al. 2005), and
is now confirmed with higher confidence and by the direct comparison
of pulsed profiles obtained from the same instrument (see Figure 1).
The lower value of the pulsed fraction appears therefore to be a
long lasting consequence of the giant flare and must be taken into
account when comparing the pulsed fluxes seen by \RXTE\ (Woods et
al. 2005) with the total fluxes measured by imaging satellites.

Also the spin-down trend of \sgr\ appears to have changed after the
flare: the four spin periods measured by \xmm\ in 2003--2004 could
be linearly fit with $\dot{P}\simeq$5.5$\times$10$^{-10}$ s
s$^{-1}$, but the period found in this last observation is smaller
than the extrapolation of this trend. It is instead consistent with
the slower spin-down rate $\dot{P}\simeq$1.8$\times$10$^{-10}$ s
s$^{-1}$, measured during the first \RXTE\ observations performed
after the giant flare (Woods et al. 2005). Together with the change
in pulsed fraction, this result suggests that a substantial
reconfiguration of the magnetosphere has occurred, very likely
related to the large amount of energy released on 2004 December 27.
In the model of magnetar's twisted magnetosphere (Thompson et al
2002) such a large scale modification is foreseen after a giant
flare, since the magnetosphere should relax into a less twisted
configuration. As a consequence, the bursting activity should
decrease and the spectrum should become softer. Only two bursts are
detected by the PN in the $\sim$25 ks observation done after the
flare, while in 2004 almost 70 bursts were detected in $\sim$70 ks,
with the PN in the same configuration. Therefore, the burst activity
had indeed dropped, as already reported in Rea et al. (2005), but
not completely stopped (see also, e.g., Palmer el al. 2005b,
Golenetskii et al. 2005). Also, the spectral softening is confirmed
by the power-law fit of the PN spectrum. However, the high quality
PN data allow us to establish that the \sgr\ spectrum is not
compatible with a simple power-law model. As already found in
pre-flare data (Mereghetti et al. 2005b), the addition of a
blackbody component gives a better fit. Since
%this component
%contributes to a rather small fraction of the \sgr\ observed flux
%(see figure 1),
%and its parameters are therefore
%strongly coupled to those of the main power-law component (contour
%plot?: kT/Rbb for pre-flare and post-flare). For this reason,
the present data do not show any significant time variability of the
blackbody parameters, we have decided to study the spectral
variability of \sgr\ assuming that the blackbody does not vary with
time. The results of this analysis are shown in Figure 2: the flux
of the power-law component increased steadily in the pre-flare
observations, when the spin-down rate was very fast and the bursting
activity became higher and higher,
%culminating with the giant flare.
but after the giant flare it dropped down to a level intermediate
between those observed in September-October 2004 and in 2003.
%and the
%source has shown other indications of a more relaxed behavior, as
%predicted by the magnetar's twisted magnetosphere model (Thompson et
%al. 2002) and already noted after the first pre-flare observations
%(Woods et al. 2005, Rea et al. 2005): slower spin-down, less bursts
%and a softer X-ray spectrum.
In contrast, the power-law photon index showed only a mild evolution
with time that does not appear to be directly related to source
activity. Therefore, if a constant blackbody component is introduced
in the spectral model, the overall softening visible by the
comparison of single power-law fits of the spectra before and after
the flare appears to be caused by the power-law intensity and not by
a variation of the photon index. This result does not fit well in
the scenario of the twisted magnetosphere relaxing after the giant
flare, since the reduction of the twist angle should cause a
decrease of the scattering depth and consequently increase the
power-law slope, if the primary radiation does not vary.

An alternative interpretation of the observed spectra can be given
without the working hypothesis that the blackbody has not varied
with time. Comparing the parameters of the best fit to the
post-flare spectrum alone with those of the previous \xmm\
observations (Mereghetti et al. 2005b), an increase in the blackbody
temperature and a decrease of the photon index emerge. This might
mean that 70 days after the giant flare, the magnetar surface (or
part of it) is hotter, as shown by the higher blackbody temperature,
but also by the harder power-law, which might be produced by the
scattering of harder seed photons\footnote{In this case, the
hardening of the primary spectrum should dominate over the
steepening caused by the reduced scattering probability, possibly
induced by the untwisting of the magnetosphere}. Also in this case,
the overall spectral softening would be caused by the larger
relative contribution of the blackbody component to the X-ray
spectrum, both due to the increased blackbody temperature (and
almost constant emitting area) and the lower power-law flux.

\begin{figure}[htbp!]
\vspace{8.5cm}
%\special{psfile=gamma.ps hoffset=30 voffset=315 hscale=55 vscale=55 angle=-90} \vspace{9.5cm}
\includegraphics{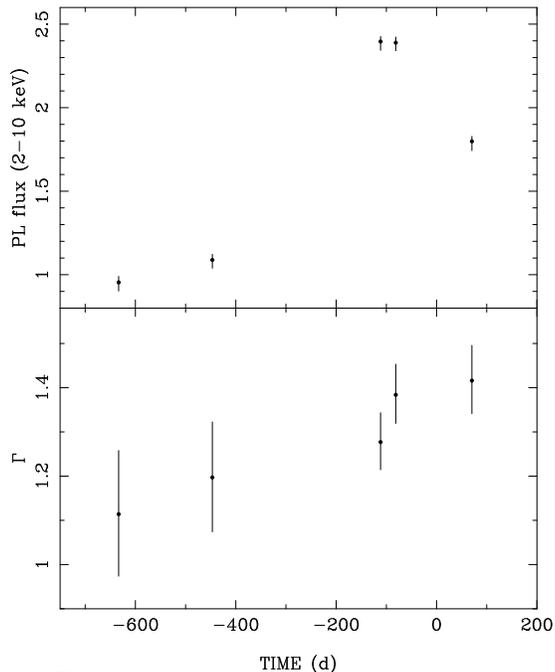} \caption{Flux of the power-law component (in
units of 10$^{-11}$ \ferg) and power-law photon index during the
five \xmm\ observations. The source spectra are fit simultaneously
using an absorbed power-law plus blackbody with only the power-law
parameters free to vary independently during the five observations.
The origin of the time axis corresponds to the giant flare of 2004
December 27.}
\end{figure}

The only other SGR that has been studied in detail after a giant
flare is \SGR\ (Kouveliotou et al. 1999, Woods et al. 2001). \RXTE\
was able to observe it rather frequently in the weeks following the
August 1998 giant flare, while the only observations performed with
imaging satellites in that period were done by \BSAX\ and
\emph{ASCA} $\sim$20 days after the giant flare. In these
observations, \SGR\ was brighter and had a softer spectrum than
during the last observations before the giant flare (Woods et al.
1999, Murakami et al. 1999). In the \BSAX\ spectrum, a blackbody
with parameters consistent with pre-flare values was also detected
(Woods et al. 1999). A hotter but rapidly cooling blackbody was
instead observed immediately after a less intense flare of \SGR\ in
April 2001 (Lenters et al. 2003) and during a period of bursting
activity of the AXP 1E~2259+586 (Woods et al. 2004); in the latter
case, the blackbody had already returned to its quiescent
temperature one day after the event, but for the much more energetic
giant flares, longer cooling times might be expected if they caused
crustal heating (Lyubarsky et al. 2002). Although the blackbody
temperatures measured by \BSAX\ 20 days after the giant flare of
\SGR\ and by \xmm\ 70 days after the much brighter flare of \sgr\
are compatible with the quiescent values, they have rather large
errors and in both cases their best-fit values are slightly higher
than in the quiescent observations.

As already mentioned, \emph{Chandra} observed \sgr\ 43 days after
the flare and measured a total flux\footnote{Note that no blackbody
component was detected in the \emph{Chandra} spectrum. However, the
data are consistent with the presence of a blackbody with the same
parameters as in the \xmm\ spectra (Rea et al. 2005), which have
better statistical quality.} $\sim$15\% higher than in the later
\xmm\ observation, which might mean that a decaying afterglow is
still contributing to the \sgr\ persistent flux at least one month
after the huge event of 2004 December 27. However these data are too
sparse to establish whether this variability is caused by a decaying
afterglow or is related to its normal activity. As a comparison, the
pulsed flux of \SGR\ measured by \RXTE\ stayed at a level higher
than usual for more than one year after the flare of 1998 August 27,
but also in that case, the simultaneous presence of moderate
bursting activity made the interpretation of this enhanced flux
uncertain (Woods et al. 2001). Therefore, the possibility that part
of the energy released in a giant flare can be stored for a long
time in the magnetar crust and then gradually emitted as thermal
radiation cannot be discarded by the present data; only high quality
spectra taken a few days after a giant flare can settle this issue.

\begin{acknowledgements}
We thank N.Schartel and the staff of the XMM-Newton Science
Operation Center for performing this Target of Opportunity
observation. This work has been partially supported by the Italian
Space Agency and by the Italian Ministry for Education, University
and Research  (grant PRIN-2004023189). NR is supported by a Marie
Curie Training Grant (MPMT-CT-2001-00245).
\end{acknowledgements}

\end{document}